\documentstyle[floats,tighten,epsf,prd,aps]{revtex}

\def\be{\begin{equation}}
\def\ee{\end{equation}}
\def\half{{1\over 2}}
\def\bea{\begin{eqnarray}}
\def\eea{\end{eqnarray}}
\def\bml{\begin{mathletters}}
\def\blea{\begin{mathletters}\begin{eqnarray}}
\def\elea{\end{eqnarray}\end{mathletters}}

\def\ba{{\bf a}}
\def\bb{{\bf b}}
\def\bx{{\bf x}}

\def\xdot{\dot{\bf x}}
\def\zhat{\hat{\bf z}}
\def\lag{{\cal L}}

\begin{document}
\draft
\wideabs{
\title{Field theory simulation of Abelian-Higgs cosmic string cusps}

\author{Ken D.\ Olum\footnote{Email address: {\tt kdo@alum.mit.edu}}
and J.\ J.\ Blanco-Pillado\footnote{Email address: {\tt
jose@cosmos2.phy.tufts.edu}}}

\address{Institute of Cosmology \\
Department of Physics and Astronomy \\
Tufts University \\
Medford, MA 02155}

\date{December 1998}

\maketitle

\begin{abstract}%
We have performed a lattice field theory simulation of cusps in
Abelian-Higgs cosmic strings.  The results are in accord with the
theory that the portion of the strings which overlaps near the cusp is
released as radiation.  The radius of the string cores which must
touch to produce the evaporation is approximately $r = 1$ in natural
units.  In general, the modifications to the string shape due to the
cusp may produce many cusps later in the evolution of a string loop,
but these later cusps will be much smaller in magnitude and more
closely resemble kinks.

\end{abstract}


}%wideabs
% Kluge for footnotes in wideabs
\def\thefootnote{\fnsymbol{footnote}}
\footnotetext[1]{Email address: {\tt kdo@alum.mit.edu}}
\footnotetext[2]{Email address: {\tt jose@cosmos2.phy.tufts.edu}}
\def\thefootnote{\arabic{footnote}}
\narrowtext
\section{Introduction}
Cosmic strings arise naturally in spontaneous symmetry breaking phase
transitions in which the vacuum manifold after the transition is not
simply connected \cite{Kibble76}. (For reviews see
\cite{Alexbook,Kibble95}).  A simple field theory which possesses such
defects is the Abelian-Higgs model,
\be\label{eqn:Lagrangian}
{\cal L} = D_\mu\bar \phi D^\mu\phi -{1\over 4} F_{\mu\nu} F^{\mu\nu}
-{\lambda\over 4}(|\phi|^2-\eta^2)^2\,,
\ee
with
\be
D_\mu\phi = (\partial_\mu -ieA_\mu)\phi\,.
\ee
We can choose units such that $\eta = 1$ and $e = 1$.  We will also
work in the ``critical coupling'' regime in which $\beta =\lambda/(2e^2)
= 1$ so that in our units $\lambda = 2$.

The fields of an infinitely long, static, straight string are
given by \cite{Nielsen73}
\bea
\phi (r) &=& e^{i\theta} f (r)\label{eqn:phi}\\ A_\theta &=& -{\alpha (r)\over r}\label{eqn:A}
\eea
where $f (0) =\alpha (0) = 0$ and $f (r)\longrightarrow 1$ and
$\alpha (r)\longrightarrow 1$ as $r\longrightarrow \infty$.  The
exact form of $f$ and $\alpha $ must be found numerically.

If the string core radius is small as compared to the radius of curvature
of a string, the action can be approximated by the Nambu action,
\be
S = -\mu\int d^2\zeta\sqrt {-\det\gamma}
\ee
where $\mu$ is the energy per unit string length, $\gamma$ is the
metric on the world sheet of the string and $\zeta^1,\zeta^ 2$ denote
the world sheet coordinates.  We can choose parameters $\sigma$ and
$t$ in such a way that the position of the string is given at any time $t$
by a function $\bx (\sigma, t)$ that satisfies
\bml\label{eqn:parameters}\bea
|\bx' (\sigma, t)| ^ 2 + |\xdot (\sigma, t)|^2 & = & 1\label{eqn:xpxd}\\
\bx' (\sigma, t)\cdot\dot\bx (\sigma, t) & = & 0 \label{eqn:xpxd2}\,.
\elea
The equation of motion is
\be
\bx'' (\sigma, t) =\ddot\bx (\sigma, t)\,,
\label{eqn:eom}
\ee
where $\bx'$ denotes differentiation with respect to $\sigma$ and
$\dot\bx$ denotes differentiation with respect to $t$.  Equation
(\ref{eqn:xpxd2}) means that the velocity of the string is always in
the plane perpendicular to the string while Eq.\ (\ref{eqn:xpxd})
parameterizes the string with constant energy per unit $\sigma$.
The general solution of Eqs.\ (\ref{eqn:parameters}) and (\ref{eqn:eom}) is
\be\label{eqn:general}
\bx (\sigma, t) =\half (\ba (\sigma - t) +\bb (\sigma + t))\,,
\ee
where $\ba$ and $\bb$ are arbitrary functions that satisfy $|\ba' | =
|\bb' | = 1$.  The functions $\ba$ and $\bb$ describe waves traveling
in the positive and negative $\sigma$ directions, respectively.

As long as the string remains smooth, its entire evolution is given by
Eq.\ (\ref{eqn:general}) and the state of the string
at all times is fixed by the functions $\ba' $ and $\bb'$.  The values
of these functions are unit vectors, so we can consider
them to trace out paths on the surface of the unit sphere.  For a
string loop, all quantities will be periodic in $\sigma$, so the paths
on the unit sphere will be closed.  Furthermore, since $\ba$ and $\bb$
are themselves periodic,
\be
\int d\sigma\ba'(\sigma) =\int d\sigma\bb'(\sigma) = 0\,,
\ee
which is to say that the center of gravity of the paths of $\ba'$ and
$\bb'$ is at the center of the sphere.  Thus, for example, $\ba'$ or
$\bb'$ could not lie entirely in a single hemisphere, and, in general,
we would expect that there would be one or more crossings between the
paths of $\ba'$ and $\bb'$.  By the same argument, there will be places
where $\ba'$ and $-\bb'$ have the same value, that is to say values of
$\sigma_a$ and $\sigma_b$ at which $\ba'(\sigma_a) +\bb'(\sigma_b) =
0$.  As a result, at time $t = (\sigma_b-\sigma_a)/2$ and position
$\sigma = (\sigma_b +\sigma_a)/2$ we will find $\bx'(\sigma, t) =
(\bb'(\sigma + t) +\ba'(\sigma -t))/ 2 = 0$ and $|\xdot (\sigma,
t)|=|(\bb'(\sigma + t) -\ba'(\sigma -t)|/2 = 1$.  Such a point is a
cusp.  The string there is momentarily moving (in the Nambu-Goto
approximation) at the speed of light.

As the string evolves toward the time of the cusp, the Nambu-Goto
approximation will break down.  Since $\bx'= 0$ at the cusp, the
segments of string point in the same direction leaving the cusp, and
so there is the potential for overlap between the segments on the two
sides.  When the string cores overlap, the topological constraints
which stabilize the string no longer operate, and the energy in the
string can be released as radiation. 
This radiation is of cosmological interest because it could
potentially lead to observable cosmic rays, and because it affects the
rate at which oscillating loops lose their energy
\cite{Branden87,Pijus89,Branden90,Branden93a,Branden93b,jjkdo98.0}.

At some point when the cusp-related evolution is complete, the
string will once again be smooth and non-overlapping, and will resume
Nambu-Goto evolution with new functions $\ba'$ and $\bb'$.  One can
consider the evolution at the time of the cusp to be a
scattering process which takes an initial state of $\ba'$ and $\bb'$
and produces a final state with new $\ba'$ and $\bb'$ and a certain
amount of outgoing radiation.

To determine the amount of radiation emitted, and the resulting
configuration of the string, we have written a lattice field theory
simulation of the Abelian-Higgs field theory.


\section{Theoretical Expectations}
\begin{figure}
\begin{center}
\leavevmode\epsfbox{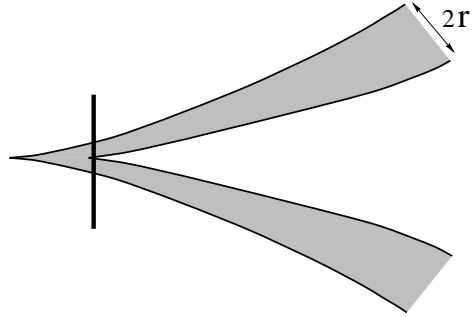}
\end{center}
\caption{Schematic shape for a cusp in a cosmic string.  The shaded region
corresponds to the core of the string.  To the left of
the vertical line, the cores overlap, and the energy could be released
as radiation.  The string appears thinner near the cusp because of
Lorentz contraction.}
\label{fig:cutoff}
\end{figure}

The usual expectation is that the string can participate in cusp
evaporation when the ``cores'' of the strings on the two sides of the
cusp are overlapping.  A parameter $r$ gives the ``core radius,''
i.e., the point at which overlap dynamics can begin, as shown in Fig.\
\ref{fig:cutoff}.  We expect this
radius to be on the order of $\eta^{-1} = 1$ in our units.

In \cite{jjkdo98.0} we showed that a generic cusp can be analyzed in a
frame in which $\bx'''_0$ is parallel to $\xdot_0$, where the
subscript $0$ denotes quantities evaluated at the cusp.  Throughout
this paper we will work in such a reference frame.

In section III of \cite{jjkdo98.0} we calculated the amount of overlap
between the two branches of a string at a cusp.  The result was that
the two segments of the string overlap from the cusp out to a distance $\sigma_c$ given
by 
\be\label{eqn:sigmac}
\sigma_c = \left(  3 r \sqrt{{{|\bx''_0|^2 + |\xdot'_0|^2}\over {|\xdot'_0 \cdot \bx''_0|^2}} 
- {1\over {|\bx''_0|^2}}}\right)^{1/2}+O(r)\,.
\ee
We expect an amount of string of total length $2\sigma_c$ to be
replaced by a short bridging string of length $2r$ or less, so
that the energy released in radiation will be approximately
$2\mu\sigma_c$.
 
In a string of cosmological size, $\sigma_c$ would be very much larger
than $r$, and the approximation in Eq.\ (\ref{eqn:sigmac}) would be
very accurate.  However, we cannot simulate a string with such a large
ratio of length scales. (Our largest simulation has $\sigma_c/r\approx
17 $.) To improve the accuracy of our prediction, we proceed as
follows.  We assume that the string has Nambu-Goto evolution until the
time of the cusp (which is confirmed by simulation, see below), and
compute the positions $\sigma_+$ and $\sigma_-$ on the two sides of
the cusp at which the (Lorentz contracted) string cores just touch.
It is possible that $\sigma_+$ and $\sigma_-$ are not uniquely defined
by this condition, since perhaps $\sigma_+$ could be moved away from
the cusp while $\sigma_-$ was moved toward the cusp to produce
different points of overlap.  In this case, we take the points which
lead to the largest emission of energy.  We then assume that the
string segment from $\sigma_-$ to $\sigma_+$ will be replaced by a
bridging string at rest between $\bx (\sigma_-)$ and $\bx (\sigma_+)$,
to arrive at an estimate of the energy release.  Since we do not know
the radius $r$ at which overlap dynamics become possible, we can
consider the simulation to provide us with a value for this radius.

\begin{figure}
\begin{center}
\leavevmode\epsfbox{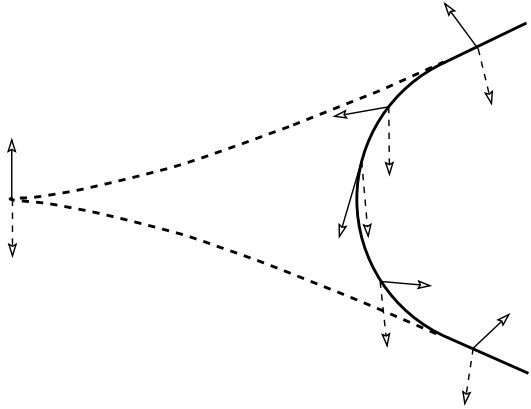}
\end{center}
\caption{Schematic of the evolution of the functions $\ba'$ (solid
arrows) and $\bb'$ (dashed arrows) before (dashed line) and after
(solid line) the cusp evaporation process.}
\label{fig:ab}
\end{figure}

\begin{figure}
\begin{center}
\leavevmode\epsfbox{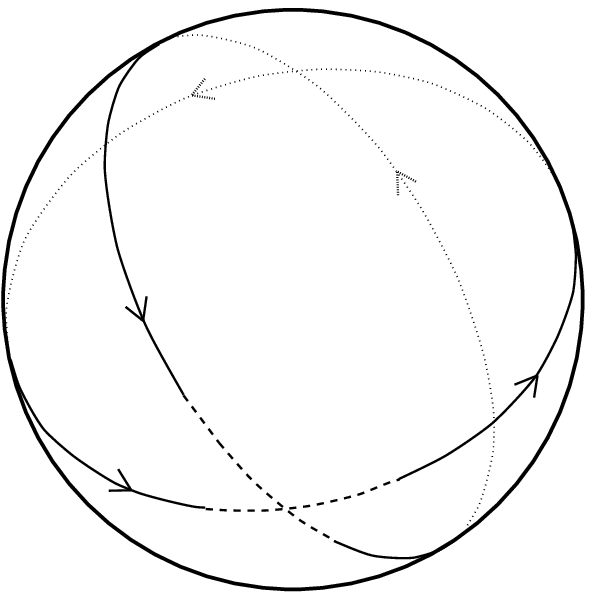}
\end{center}
\caption{The paths of $\ba'$ and $-\bb'$ before the cusp.  The dashed
segments are the parts that will be affected by the cusp.  The dotted
curves are the paths around the back of the sphere.} 
\label{fig:before}
\begin{center}
\leavevmode\epsfbox{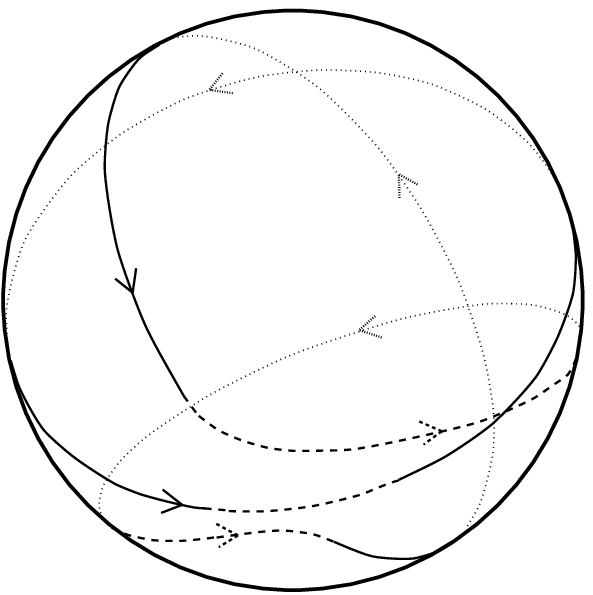}
\end{center}
\caption{The paths of $\ba'$ and $-\bb'$ after the cusp.  One of the
paths has been broken and reconnected around the back of the sphere.}
\label{fig:after}
\end{figure}

How should the paths of $\ba'$ and $\bb'$ be affected by the
evaporation of the cusp?  Near the cusp, $\ba'(\sigma)\sim\ba'_0 =
-\xdot_0$ and $\bb'(\sigma)\sim\bb'_0 =\xdot_0$.  Outside the region
of overlap, they will retain these values, but inside that region we
will have a bridging string.  The bridging string will point in the
direction of $\bx'''_0$ because that is the direction in which the two
arms of the cusp spread apart.  Since the motion of the string near
the cusp is approximately $\xdot_0$, which is parallel to $\bx'''_0$,
the bridging string will be created approximately stationary.  Thus on
the bridge, $\ba'(\sigma) =\bb'(\sigma)\propto\bx'''_0$ with a
positive constant of proportionality.  There are thus two cases,
depending on whether $\bx'''_0$ is parallel or antiparallel to
$\xdot_0$.  If they are parallel, then $\bb'(\sigma)$ on the bridge
will have roughly the same value that it has on nearby unaffected
portions of the string, but $\ba'(\sigma)$ on the bridge will have
nearly the opposite value as it does nearby, as shown in Fig.\
\ref{fig:ab}.
Thus in this case we expect the path of $\ba'(\sigma)$ to
be broken near the point of crossing and reconnected around the
opposite side of the unit sphere.  In the case that $\bx'''_0$
and $\xdot_0$ are antiparallel, we will have the reverse behavior.
The path of $\ba'(\sigma)$ will be pretty much unaffected, while
$\bb'(\sigma)$ will be broken and reconnected. This effect of the 
cusp on the evolution of the functions $\ba'$ and $\bb'$ on the unit
sphere is shown in Figs.\ \ref{fig:before} and \ref{fig:after}.

\section{Simulation}
\subsection{Lattice Action}
To put Eq.\ (\ref{eqn:Lagrangian}) on a lattice, we have used a
somewhat different approach from those that have been used in the past
\cite{Moriarty88,Hind98,Shellard98}.  We put the system on a lattice
in both space and time, with the scalar field $\phi$ stored on the
vertices of the lattice, the vector field $A_\mu$ on the links, and
the field strength $F_{\mu\nu}$ on the faces.  The action can be
specified as follows.

Let $s$ denote a lattice site, $l$ a directed link from site
$l_-$ to site $l_+$, and $f$ a directed face made of links
$f_1$, $f_2$, $f_3$, and $f_4$.  Let $\sum_s$ denote summation over
all sites, $\sum_l$ summation over all links, counting each link once,
and $\sum_f$ summation over all faces, counting each face once.  Let
$\sum_{l|l_+ = s}$ denote summation over all links that end in site
$s$ and $\sum_{f|f_1 = l}$ denote summation over all faces that border
link $l$.  Then the lattice action is
\be\label{eqn:action}
S =\Delta x^3\Delta t\left(\sum_s\lag_s +\sum_l\lag_l
+\sum_f\lag_f\right)
\ee
where
\bea
\lag_s &=& -V (\phi_s) = -{1\over 4}\lambda\left
(|\phi_s|^2-\eta^2\right)^2\\ \lag_l &=& g_lD_l\bar\phi D_l\phi\\
\lag_f&=&-\half g_fF_f^2\,.
\eea
Here $\Delta x$ and $\Delta t$ are the spatial and temporal lattice
spacing; $g_l$ is the metric, $g_l = +1$ if $l$ is timelike, and $g_l
= -1$ if $l$ is spacelike; and $g_f = g_{f_1} g_{f_2}$, i.e., $g_f =
+1$ if all sides of $f$ are spacelike, and $g_f = -1$ if $f$ has a
timelike side.
The discrete covariant derivative is given by
\blea
D_l\phi &=& \Delta_l (\phi_{l_+} -e^{i\Delta_leA_l}\phi_{l_-})\\
D_l\bar\phi& = &\Delta_l (\bar\phi_{l_+} -e^{-i\Delta_leA_l}\bar\phi_{l_-})
\elea
where $\Delta_l$ denotes $\Delta x$ if $l$ is spacelike or $\Delta t$
if $l$ is timelike.  The field strength is given by
\be
F_f =\Delta_{f_1}^{-1}\Delta_{f_2}^{-1}\left(\Delta_{f_1} A_{f_1}+
\Delta_{f_2} A_{f_2}+\Delta_{f_3} A_{f_3}+\Delta_{f_4} A_{f_4}\right)\,.
\ee

The action of Eq.\ (\ref{eqn:action}) is invariant under a discrete
gauge transformation given by a number $\Lambda_s$ at each lattice
site, 
\blea \phi_s&\longrightarrow &e^{i\Lambda_s}\phi_s\\
A_l&\longrightarrow&A_l +\Delta_l^{-1} (\Lambda_{l_+} -\Lambda_{l_-})\,.
\elea
We will work in temporal gauge, $A_l = 0$ for all timelike links
$l$, which greatly simplifies the calculation.

The equations of motion resulting from Eq.\ (\ref{eqn:action}) are
\bml\label{eqn:lattice-eom}\bea 
\sum_{l|l_+ = s} g_l\Delta_l^{-2} (\bar\phi_{l_-}
e^{-i\Delta_leA_l} -\bar\phi_s) +{\partial V\over\partial\phi_s} &=&
0\\ \sum_{f|f_1 = l}\Delta_{f_2}^{-1} g_fF_f &=& j_l
\elea
where the link current is
\be
j_l = ie\Delta_l^{-1} g_l (e^{-ie\Delta_lA_l}\bar\phi_{l_-}\phi_{l_+} -
e^{ie\Delta_lA_l}\bar\phi_{l_+}\phi_{l_-})\,.
\ee

The stability of our finite difference scheme requires the Courant
condition, $\Delta t / \Delta x < 1/\sqrt 3 \approx 0.58$ \cite{NR},
for three spatial dimensions. In fact, ratios slightly less than this
are still unstable in our code, presumably due to numerical error.  To
be safe, we have used $\Delta t / \Delta x = 0.5$.  To verify that
this condition is sufficient, we have checked that our simulations
conserve energy over long periods of time.

We have used reflecting boundary conditions (fixed field values) on
the top and the bottom faces of our lattice box, and first order
absorbing boundary conditions \cite{Absorbing77,Absorbing90} on the
lateral faces.  The absorbing conditions prevent the radiation emitted
at the cusp from reflecting off the walls and interacting with the
later evolution of the string.

\subsection {Initial Conditions}

To simulate cusp formation we must have an initial field configuration
which will evolve into a cusp.  This can be done exactly, using a
result of Vachaspati et al.\ \cite{Tanmay90} that a traveling wave
along a straight string is an exact solution of the field theory
equations of motion.  In other words, an arbitrarily shaped wiggle that
propagates at the speed of light along a straight string will not
radiate and will retain its form.

These wiggles can be seen also in terms of functions $\ba'$ and $\bb'$
in the Nambu-Goto approximation.  For example, a straight part of the
string in the $\zhat$ direction can be written in terms of $\ba' =
\zhat$ and $\bb' = \zhat$.  Introducing a wiggle traveling in one
direction along the string will modify one of the functions $\ba'$
or $\bb'$ depending on the direction of the movement of the wiggle,
keeping the other function fixed.

Using this method we can obtain the Nambu-Goto description for two
wiggles traveling toward each other along a straight string.  Then by
using \cite{Tanmay90} we can write down the expression for the fields
in terms of the functions $\ba'$ and $\bb'$ and the static string
fields given by Eqs.\ (\ref{eqn:phi}) and (\ref{eqn:A}).  By
calculating the values for the fields when the two wiggles are still
separated by a portion of straight string we ensure that our initial
conditions are exact, so we do not have to use a relaxation procedure
as is necessary in other field theory simulation schemes
\cite{Moriarty88,Hind98,Shellard98}.

We design the two colliding wiggles such that they will produce a cusp
in the Nambu-Goto evolution and use them as our initial conditions in
the field theory code.  One could imagine that the string would evolve
into a different configuration due to interaction before the time of
the cusp, but as we will show later, that does not occur: the string
core in the field theory simulation follows the path that the
Nambu-Goto evolution predicts until immediately before the cusp.

As described earlier, the cusp is characterized in the Nambu-Goto
approximation by the intersection of $\ba'$ and $-\bb'$ on the unit
sphere.  Far outside the area of interaction, the string will be
straight with $\ba'$ at the north pole of the unit sphere, and $-\bb'$
at the south pole. Thus we are looking for functions $\ba'$ and $\bb'$
which make loops on the sphere, starting and finishing at the poles
and crossing at some point of the sphere.

Two closed loops cannot intersect in just a single point, unless it is
a point of tangency, which does not give rise to a generic cusp.  But
we want to obtain information about the energy radiated by just one
cusp, so we do not want to have any interference in the dynamics due
to the formation of a second one.  (A realistic cosmological string
would have multiple cusps, but they would be very well separated.)  To
prevent this interference we will design the two wiggles so that the first cusp
will alter the string in such a way that the second one will not
occur.  We will choose our parameters so that $\bx'''_0\cdot\xdot_0 >
0$, so we expect $\ba'$ to be broken and reconnected, while the path
of $\bb'$ will not be greatly affected.  Thus we will arrange for
$\bb'$ to cross $\ba'$ twice at the same point, so that when $\ba'$ is
reconnected to eliminate the first crossing, the second crossing will be
eliminated as well.  We will choose a simple path for $\bb'$ that
starts at the pole, runs along a meridian, crosses $\ba'$ at the
equator, and continues for short distance before turning around and
returning along the same path.

We must also arrange that the magnitudes of $\ba''_0$ and $\bb''_0$ are
different to avoid a pathological situation which leads to extra
energy emission, as discussed in \cite{jjkdo98.0}.  We also want to
work in the canonical frame where $\bx'''_0$ is parallel to $\xdot_0$,
so we require that $\ba'''_0 +\bb'''_0$ is perpendicular to $\ba''_0$
and to $\bb''_0$.

We will start by specifying our desired values for
$\ba''_0$ and $\ba'''_0$.  As much as possible we would like other
derivatives not to contribute near the cusp, so we will set
\be
\ba'_{\text{unnormalized}} (\sigma) = \ba'_0 +\sigma\ba''_0
+{\sigma^2\over 2}\ba'''_0
\ee
and then
\be
\ba' (\sigma) = {\ba'_{\text{unnormalized}}(\sigma)\over
|\ba'_{\text{unnormalized}}(\sigma) |}\,.
\ee
We will use this form for $\sigma \in [-\sigma_1,\sigma_1]$, where
$\sigma_1$ is chosen to be large enough that changes in $\ba'$ outside
this range will not have too much effect on the cusp dynamics.

Outside of this range, we just have to connect $\ba'$ 
to the north pole of the unit sphere.  We do this by choosing paths
of  circles on the sphere which will smoothly interpolate
between $\ba'( -\sigma_1)$ and $\ba'(\sigma_1)$ and the pole.

\begin{figure}
\begin{center}
\leavevmode\epsfbox{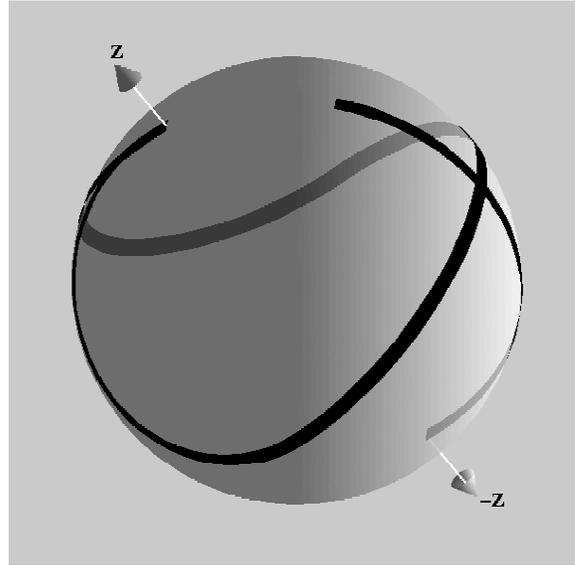}
\end{center}
\caption{Initial conditions for the functions $\ba'$ and $-\bb'$, on the
unit sphere.} 
\label{fig:ini}
\end{figure}

The result of this design for $\ba'$ and $-\bb'$ is shown in Fig.\
\ref{fig:ini}.

\begin{figure}
\begin{center}
\leavevmode\epsfbox{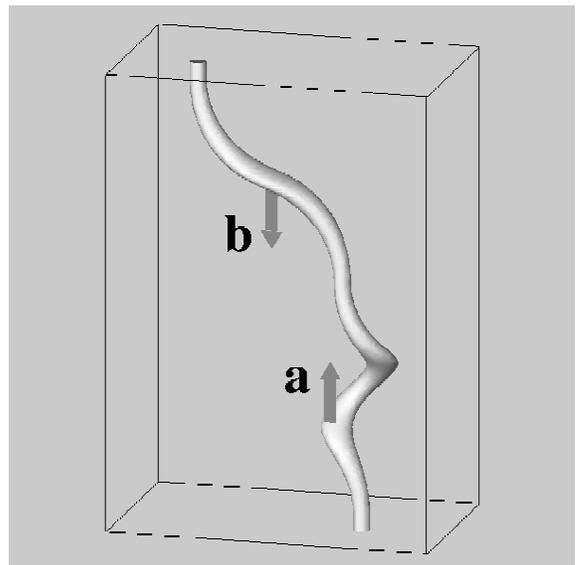}
\end{center}
\caption{The initial conditions.  The surface shown is a surface of
constant energy density surrounding the string core.} 
\label{fig:initial-string}
\end{figure}

To produce the initial field configuration, we offset the wiggles
produced by $\ba'$ and $\bb'$, so that one is at the top of the
lattice moving downward and other is at the bottom of the lattice moving
upward.  The actual shape of the string can be seen in Fig.\
\ref{fig:initial-string}

When the two wiggles combine, we will look at the shape of the string
at the time of the cusp in the Nambu-Goto approximation, and define
the ``length scale'' of the cusp for the sake of discussion as
\be\label{eqn:scale}
L = {1\over|\bx''_0|}\,.
\ee

\subsection{Recovering the energy and form of the string.}
	
In order to obtain the amount of energy released 
in the cusp evaporation, we implemented a method to recover, 
out of the field theory simulation, the total amount of energy
left in the string.

We first identify the plaquettes of the lattice that the string goes
through, by looking for places where the phase of the complex scalar
field on the four corners of the plaquette wraps around the origin in
the complex plane.  As long as two strings do not go through the same
lattice cube, this technique always allows the string to be traced
from one face to the next.  We take the path of the string to go
through the centers of these plaquettes.  (We have tried some
algorithms for improving the path of the string by closer examination
of the field values at the corners of the plaquettes, but they did not
produce significantly better results.)

This technique produces a discrete approximation to the string path,
which we then smooth by a Gaussian convolution.  The end result is a
list of smoothed strings at different times, parameterized by an
arbitrary parameter $\lambda$.  We will denote these string positions
by $\bx (\lambda, t)$.

Near the time and position of the cusp, the dynamics are not well
described by the motion of a string, but rather involve the complete
field configuration.  One can still recover a path of faces around
which the field phase winds, but the different positions of this path
at different times do not represent motion, but rather the unraveling
of a field configuration.  However, since we are only interested in the
difference between the energy in the string before and after the cusp,
we will just apply this method at times well before or well after the
cusp where we are certain that the Nambu approximation is a faithful
description of the string.

To recover the energy of the string we first calculate 
$\xdot (\lambda, t)$ by computing
\be
\xdot (\lambda, t) = {{\bx(\lambda,t) - \bx (\lambda', t-N\Delta t)}
\over N {\Delta t}}
\ee
where $N$ is a number of lattice time steps over which we average to
reduce the effect of errors in recovering the string position, and
$\lambda'$ is determined by the condition that $\left(\bx(\lambda,t)
- \bx (\lambda', t-N\Delta t)\right)$ be perpendicular to
$d\bx(\lambda,t)/ d \lambda$.

Once we have calculated the velocity we can obtain the relation
between $\lambda$ and the usual parameter $\sigma$ by using Eq.\
(\ref{eqn:xpxd}),
\be 
\left|{{d\bx(\lambda,t)}\over {d\lambda}}\right|{{d \lambda}\over {d\sigma}}  = \sqrt {1-  |\xdot (\lambda,t)|^2}
\ee
so
\be 
d\sigma  = {{d\lambda} \over { \sqrt {1-  |\xdot (\lambda,t)|^2}}} \left|{{d\bx(\lambda,t)}\over {d\lambda}}\right|\,.
\ee

The total length of string is just the integral of the equation above
and so the energy is
\be
E_{string} = \mu \int d\sigma\,.
\ee

Once we have obtained the vectors $\bx'$ and $\xdot$ for the string,
we can get the functions, $\ba'(\sigma,t)$, and $\bb'(\sigma,t)$ from
the relations
\blea
\ba'(\sigma, t)  &=& \bx'(\sigma,t) - \xdot(\sigma,t)\\
\bb'(\sigma, t) &=& \bx'(\sigma,t) + \xdot(\sigma,t)
\elea
and study the change of these functions due to the cusp dynamics.

\section{Results}

We ran the simulation code for cusp scales (as defined by Eq.\
(\ref{eqn:scale})) from $L\approx 5$ up to $L\approx 34$.  For sizes
up to $L\approx 17$ we used $\Delta x = 0.25$ or less, which gives a
quite accurate simulation.  For larger sizes we were forced by memory
constraints to use larger values of $\Delta x$ up to $\Delta x =
0.425$.  These larger values produce somewhat decreased accuracy,
especially in those parts of the string which are rapidly moving.
They also reduce the accuracy with which the position of the string
and thus the energy emission can be recovered from the simulation.
However, we feel that these simulations are still sufficiently
accurate to draw some general conclusions about the cusp evaporation
process.

The code was written in Lisp and executed on a Digital Alpha
processor.  The largest lattices required about 700 MB of memory and
one week of CPU time.

We first run the code letting only one wiggle evolve on the straight
string, setting $\bb'=\zhat$.  This experiment confirms Vachaspati's
result \cite{Tanmay90}, showing the traveling wave on the string
without any radiation. This is also a check for the code, since it
shows stability and energy conservation for the evolution with the
boundary conditions specified above.

\begin{figure}
\begin{center}
\leavevmode\epsfbox{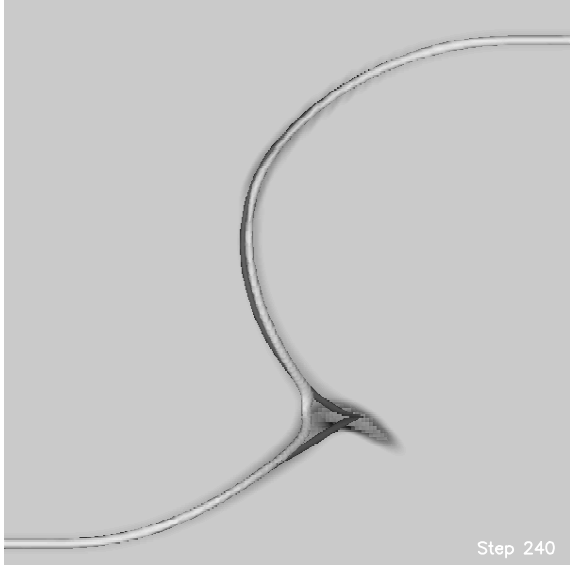}
\end{center}
\caption{String evolution at the cusp for $L\approx 17$.  The core of
the string in the field theory simulation is shown light, and the
Nambu-Goto prediction dark.  The ``fog'' shows areas of high energy
density.  The string core has collapsed away from the point of the
cusp, leaving the energy behind in radiation.}
\label{fig:cusp}
\end{figure}

\begin{figure}
\begin{center}
\leavevmode\epsfbox{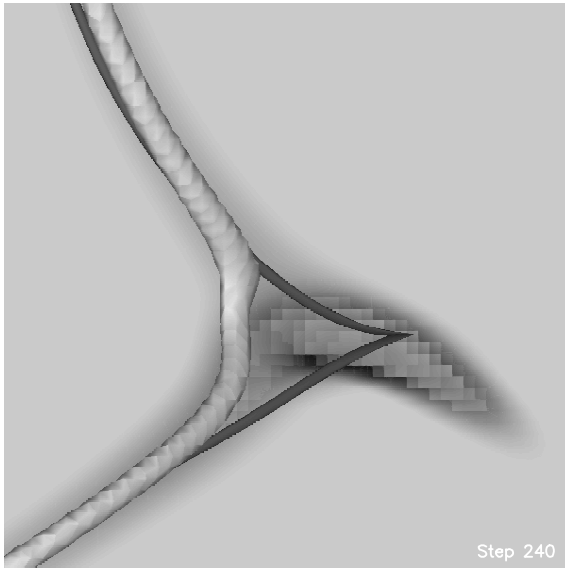}
\end{center}
\caption{Close-up view of Fig.\ \ref{fig:cusp}}
\label{fig:cusp-big}
\end{figure}

\begin{figure}
\begin{center}
\leavevmode\epsfbox{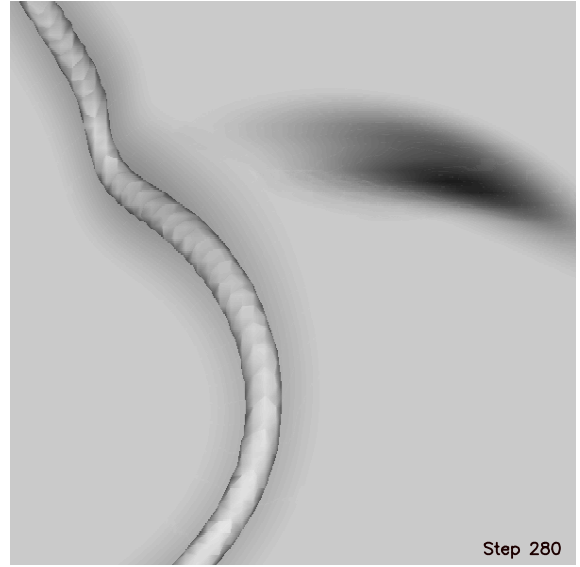}
\end{center}
\caption{Close-up view of the core of the string and the radiation after the
cusp evaporation. The cloud of radiated energy travels in an expanding shell.}
\label{fig:cusp-after}
\end{figure}

With the two colliding wiggles, the string forms a cusp as expected.
Energy is conserved to within 1\% until the radiation emitted at the
cusp begins to be absorbed at the boundary.  Figures \ref{fig:cusp}
and \ref{fig:cusp-big}
show the string shortly after the moment at which the Nambu-Goto
approximation predicts the cusp.  Except near the cusp, the Nambu-Goto
evolution has been followed accurately.  But near the cusp, the core
of the string has collapsed away from the area where the cusp energy
is stored and has released this energy in the form of radiation.  This
is what one would expect from the overlap model.  At some point near
the cusp the strings are close enough that the fields can untangle,
and the cusp beyond this point is replaced by a bridging string.
After this time, the released radiation travels freely in an expanding
shell as shown in Fig.\ \ref{fig:cusp-after}.

\begin{figure}
\begin{center}
\leavevmode\epsfbox{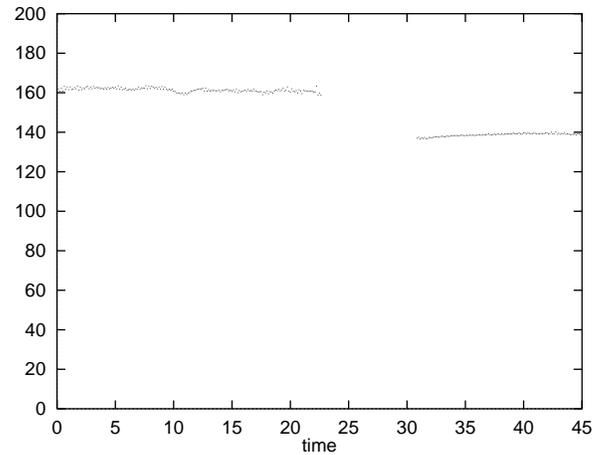}
\end{center}
\caption{Recovery of the total amount of $\sigma$ on the string during
a cusp evaporation, for $L\approx 17$
.  Points at left are before the cusp; points at
right are those after cusp evaporation. Near the time of the cusp,
$\sigma$ cannot be recovered.}
\label{fig:recovery}
\end{figure}

For each simulation run, we recover the shape of the string, the form
of $\ba'$ and $\bb'$, and the total length of string at each time
step.  Figure \ref{fig:recovery}
shows a typical graph of the recovered points.  There is significant
variation in the individual points, both random variation due to
the way in which the string core cuts through the lattice, and
systematic variation due to inaccuracies in recovering the energy that
depend on the string shape.  However, there is a clear difference
between the string length before and after the time of the cusp.  (At
times very near the cusp the energy cannot be recovered so there are no
corresponding points on the plot.)

\begin{figure}
\begin{center}
\leavevmode\epsfbox{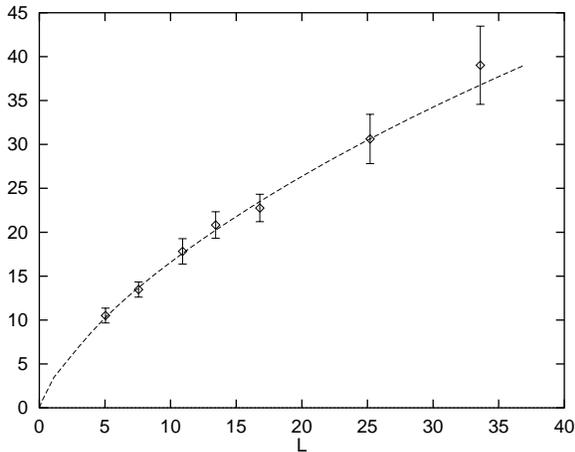}
\end{center}
\caption{Amount of $\sigma$  as a function of the length scale $L$. The 
dashed line corresponds to the theoretical prediction assuming $r= 1.15$.}
\label{fig:r}
\end{figure}

\begin{figure}
\begin{center}
\leavevmode\epsfbox{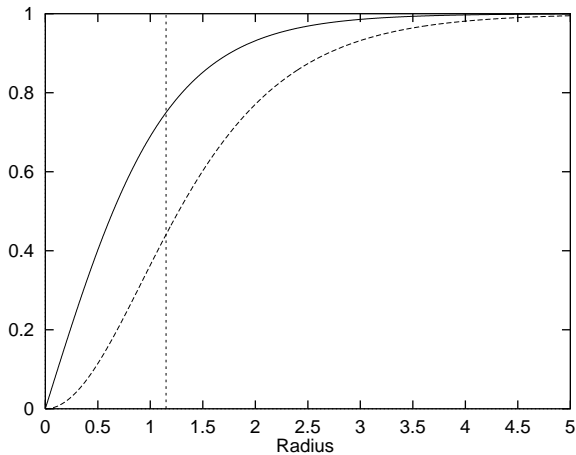}
\end{center}
\caption{Functions $f$ (solid curve) and $\alpha$ (dashed curve) for
the static straight string. The vertical line indicates the best-fit
value of $r$ from the field theory simulation.}
\label{fig:shape}
\end{figure}

In Fig.\ \ref{fig:r}
we plot the string length released by the cusp and compare with
theoretical predictions.  The error bars in Fig.\ \ref{fig:r} are a
combination of the variation in the actual recovered energy with an
estimate of the systematic errors due to the recovery process.  As
shown, the results are in accord with the overlap model with overlap
radius $r\approx 1.15$.  Figure \ref{fig:shape}
shows this value of $r$
as compared to the shape of the string profile.  One would expect the
overlap dynamics to begin at a radius where the fields have values
intermediate between the core values and those at large distances, and
the result seems quite reasonable in this regard.

\begin{figure}
\begin{center}
\leavevmode\epsfbox{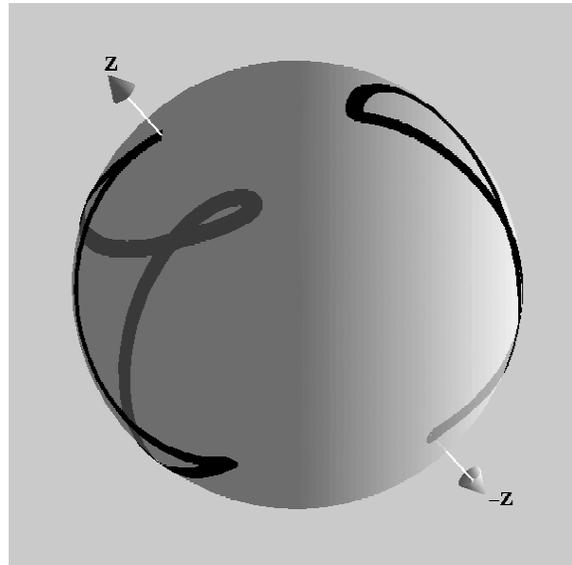}
\end{center}
\caption{Recovered functions $\ba'$ and $-\bb'$ after the cusp, for 
$L \approx 17$.}
\label{fig:ab-after}
\end{figure}

The recovered $\ba'$ and $-\bb'$ shortly after the cusp are shown in
Fig.\ \ref{fig:ab-after}.
By comparison with Fig.\ \ref{fig:ini}, we can see that the
path of $-\bb'$ is roughly the same before and after the cusp, but the
path of $\ba'$ has been greatly altered.  Far from the cusp it is the
same, but in the region of interaction, it has been reconnected
around the other side of the sphere from the original crossing point.
This reconnection, as planned, prevents the occurrence of a second
cusp.

\section{Discussion}

We have simulated a cosmic string cusp in Abelian-Higgs field theory
on the lattice.  Our results are in accord with the model that the
evolution is accurately given by the Nambu-Goto equations of motion
until nearly the time of the cusp.  The effect of the cusp is to
release an amount of energy which is well approximated by the overlap
model with $r\approx 1.15$.

As discussed in \cite{jjkdo98.0}, for a string of cosmological size,
this model gives an energy emission of order $\mu\sqrt{rL}$,
where $L$ is the typical length scale.  This is smaller than previous
estimates \cite{Branden87,Pijus89,Branden90,Branden93a,Branden93b} by
a factor of $(r/L)^{1/6}$ and would not lead to observable cosmic
rays.

The effect of the evolution through the cusp is to disconnect the path
of either $\ba'$ and $-\bb'$ and reconnect it around the other side of
the unit sphere.  We will suppose without loss of generality that
$\bx_0'''\cdot\xdot_0 > 0$ so that it is $\ba'$ that is reconnected,
while $-\bb'$ merely has a section deleted and forms a small kink.  It
is likely that the reconnected $\ba'$ will have many crossings with
$-\bb'$ and thus might lead to many cusps in the future.  However, the
total amount of string length in which $\ba'$ loops around the sphere
is quite small, on the order of $r$.  Thus the remaining feature
involves a bending of the string in about a string radius, and is thus
more like a kink than a region of string which would produce a cusp.
Any cusps in which it is involved will have even smaller emission of
radiation than the original cusp. 

This result could perhaps be modified by gravitational radiation,
which smoothes out small-scale features \cite{Quash}.  Gravitational
back-reaction might smooth out the region where $\ba'$ varies rapidly
and restore it to its original crossing with $-\bb'$.  This would
produce another cusp at the same position in a subsequent oscillation.
However, the region of overlap contains much less energy than before,
because most of it has been emitted, so that $\ba'$ and $-\bb'$ cross
this region in a small amount of $\sigma$.  Subsequent gravitational
effects might increase this energy, but it will not grow as large as
in the original cusp, and so will not lead to observable radiation, as
discussed above.

\section{Acknowledgements}

We would like to thank Inyong Cho for providing us with a static
string profile, and Xavier Siemens and Alex Vilenkin for helpful
conversations.  This work was supported in part by funding provided by
the National Science Foundation.  J. J. B. P. is supported in part by
the Fundaci\'on Pedro Barrie de la Maza.

%\bibliography{cosmic-string}
%\bibliographystyle{prsty}

\end{document}